\documentclass[letterpaper]{sig-alternate}
\usepackage{verbatim}
\usepackage{url}
\usepackage[english]{babel}
\begin{document}

\setlength{\pdfpageheight}{\paperheight}
\setlength{\pdfpagewidth}{\paperwidth}

\pdfpagewidth=8.5in
\pdfpageheight=11in

\title{Offloadable Apps using SmartDiet: Towards an analysis toolkit for mobile application developers}
%
%
%
%
%

\numberofauthors{2} 
%
\author{
%
%
\alignauthor
  Aki Saarinen, Matti Siekkinen, Yu Xiao, Jukka K. Nurminen, Matti Kemppainen\\
  \affaddr{Aalto University School of Science, Finland}\\
  \email{aki@akisaarinen.fi, \{matti.siekkinen, yu.xiao, jukka.k.nurminen\}@aalto.fi, matti.kemppainen@iki.fi}
\and
\alignauthor
  Pan Hui \\
   \affaddr{Deutsche Telekom Labs}\\
   \affaddr{Berlin, Germany}\\
   \email{pan.hui@telekom.de}
}
\date{7 October 2011}

\maketitle
\begin{abstract}
Offloading work to cloud is one of the proposed solutions for increasing the battery life of mobile devices. Most prior research has focused on computation-intensive applications, even though such applications are not the most popular ones. In this paper, we first study the feasibility of method-level offloading in network-intensive applications, using an open source Twitter client as an example. Our key observation is that implementing offloading transparently to the developer is difficult: various constraints heavily limit the offloading possibilities, and estimation of the potential benefit is challenging. We then propose a toolkit, SmartDiet
\footnote{SmartDiet is available under an open source license at \url{https://github.com/akisaarinen/smartdiet}}, to assist mobile application developers in creating code which is suitable for energy-efficient offloading. SmartDiet provides fine-grained offloading constraint identification and energy usage analysis for Android applications. In addition to outlining the overall functionality of the toolkit, we study some of its key mechanisms and identify the remaining challenges.
\end{abstract}

\section{Introduction}

A seemingly straightforward way to save energy of a mobile device is
to offload work to a more powerful machine. Several frameworks have
been proposed for \textit{computation offloading} including
MAUI\cite{cuervo10maui}, Cuckoo\cite{kemp10cuckoo},
CloneCloud\cite{chun11clonecloud}, and
ThinkAir\cite{sokol11thinkair}. They support method-level migration of
software execution and try to trade the energy spent in migration for
the savings gained from the reduction of local computation.

These frameworks share two common characteristics. One is that they fail to provide any tools or guidelines to help offload existing programs. The other is that they
focus only on heavy computation offloading. We challenge these design
decisions on the basis of our findings and argue that an alternative path should be taken for two reasons. First, automated offloading using existing frameworks has
certain limitations. To make existing applications offloadable,
modifications to source code is often needed and the process is
laborious. Second, many of the popular mobile applications, such as
Facebook and Twitter, require almost no computation, but a lot of
communication. Intuitively, offloading traffic seems to make no sense since content needs to reach the mobile device anyway. However, energy savings gained through \textit{communication
  offloading} are possible because communication often contains costly signaling traffic \cite{signals103g}, part of which can be suppressed. Furthermore, packet
interval patterns and throughput, both of which have
a significant impact on communication cost\cite{yu10eenergy}, can be optimized by means of offloading.

To address these issues, we propose a toolkit, namely SmartDiet, which helps
application developers to study the offloadability of existing applications and, in turn, to implement offloading in an energy-efficient
manner. The toolkit identifies potential trouble spots in Android applications' Java source
code. It also estimates current energy usage at a fine-grained level in order to provide estimates of how much could be saved by offloading, and analyzes the software structure to identify opportunities for further savings. The following three points summarize our contributions.

1) We take the first look into the feasibility of applying
communication offloading to mobile applications, and analyze the
factors that may limit the energy savings. We use an open source
Twitter client to exemplify the associated issues.

2) We propose a toolkit, SmartDiet, for identifying constraints from existing program code. It guides programmers to improve the design and implementation of programs so that the existing  offloading frameworks can be better utilized.

3) SmartDiet provides a novel way to estimate the savings in
communication energy cost. Our method is more fine-grained than those used in
current offloading frameworks, and, moreover, takes traffic patterns and
power saving modes into account. This part of the toolkit could also be
integrated into existing frameworks to enhance the accuracy of their
runtime cost estimation.

We motivate the need for SmartDiet in Section \ref{sec:use-case} by examining an offloading use case . We then describe our vision and current state of the toolkit in Section \ref{sec:toolkit}. Section \ref{sec:challenges} discusses the remaining challenges and future work before conclusions.

\section{Related work}

Two main approaches have been suggested for mobile application
offloading. MAUI\cite{cuervo10maui},
Cuckoo\cite{kemp10cuckoo} and ThinkAir\cite{sokol11thinkair} implement a framework on top of
the existing runtime system. These
three systems are fairly easy to deploy because they only require access to the program source code, and they do not need any special support from the operating system. The second approach, used by
CloneCloud\cite{chun11clonecloud}, is to modify the underlying virtual
machine or operating system in order to implement richer mechanisms for
offloading. CloneCloud is a fully automated system and does not
require having the source code of the program, because it
works directly on bytecode. We
claim that the developer should participate in the offloading
process and therefore focus on the first approach.

Specific solutions, such as Catnap\cite{dogar10catnap}, have been proposed in the literature for reducing the communication energy cost by
applying a proxy or middlebox approach. However, these solutions will provide energy savings only for the communication part of the
program, whereas offloading simultaneously provides savings in computational costs. Furthermore, since systems such as Catnap do not execute
application logic at the proxy, all traffic must eventually reach the mobile device. With smart offloading, some part of
the traffic (e.g., signaling) might never need to reach the
mobile device, because the offloaded part of the program handles it
directly.

\section{Use case: AndTweet}
\label{sec:use-case}

To gain experience in offloading existing network-intensive programs, we offloaded the communication part of a typical such application using ThinkAir. The application was AndTweet, an open source Android Twitter client which has been downloaded from the Android Market site over 5000 times. Over the course of this study, we had to overcome a number of obstracles in our efforts to remotely execute parts of the application. After resolving these issues, we measured the energy consumption of offloaded and non-modified versions of AndTweet. We show that, although somewhat promising, the results vary considerably depending on circumstances. Our experiences provide the motivation for our toolkit which we detail in the remaining sections.

\subsection{Offloading setup}

We used ThinkAir offloading system\cite{sokol11thinkair}, which
allows us to offload a selected set of methods to a remote server. We
used a Google Nexus One with Android 2.3 as the local device and a
virtual machine running the Android x86 port as the remote execution
platform. We chose the offloaded methods manually and disabled all the
dynamic decision making features of ThinkAir.

Upon starting an application, an execution controller from ThinkAir is
instantiated and an application image is sent to a remote
server. Whenever a method marked as offloadable is executed,
the controller transfers the execution to the server. When a method is executed remotely, the client serializes the class instance of the called method and all of its arguments using Java's
serialization APIs. The server de-serializes these objects,
invokes the specified method, and sends the returned value (or
exception) back to the client. 

\subsection{Problems encountered during offloading AndTweet}
\label{sec:constraints-of-remote-execution}

We offloaded as many parts of AndTweet as possible which communicate with the Twitter backend. Our goal was to reduce the signaling traffic and hence save energy. We identified events in the user interface (UI) which trigger network requests and proceeded to search for methods whose execution could be migrated to the remote server, starting from the method which handles the UI event. We encountered several challenges, related to 1) methods accessing local hardware, 2) methods whose migration to remote server was not possible without modifications and 3) methods accessing state that is not correctly synchronized between the device and the server and therefore cause unexpected behavior.

A remotely executed method cannot interact with the UI or other hardware resources, because these resources only exist at the client. AndTweet mixes application logic and handling of UI in many of its methods that are interesting in terms of offloading. These pieces of application logic, which might be offloadable on their own, are tied to the device. To overcome the restrictions of UI interactions, we tried to offload the methods that UI handling methods directly invoke. Here we could find Twitter-specific abstractions, such as friend timeline, which is a list of timestamped messages from the people you follow in Twitter. This set of methods, however, contained another category of issues.

In order to migrate the execution of a method to a remote server, the offloading system must transfer the related dependencies over the network. Section \ref{sec:constraint-identification-tool} discusses migration further, but in case of ThinkAir this sets the requirement that the encapsulating class of an offloaded method must be serializable using Java's serialization APIs. In addition, they must not access any state outside the serialized context, because the changes are not automatically synchronized either to the remote server or back. The classes encapsulating Twitter timelines contained instances of non-serializable classes. AndTweet stores some of its internal state, for example cryptographical tokens required for authentication, using an Android standard library class SharedPreferences, which is not serializable. Similarly, HttpClient, the de-facto class for doing HTTP communication in an Android application, is also not serializable. 

Methods using HttpClient were fixed by instantiating a new HttpClient in the remote server. Existing instances now render migration unnecessary to either direction. SharedPreferences, on the other hand, was more problematic, because behind the scenes it uses the local file system to save its state. Instantiating a new SharedPreferences in the remote server, and blindly using it would result in different states between the client and the server. Therefore, we needed to manually implement a mechanism which synchronizes the remote SharedPreferences before and after a method is executed remotely. Otherwise the cryptographical tokens were out-of-sync between the device and the remote execution server. The problems with state synchronization are difficult to resolve because in the worst case no errors are reported. Instead, the method just does something wrong and the program, not to mention the developer, is unaware about it.

Detecting and fixing the aforementioned problems was especially difficult in AndTweet because classes and methods had a large number of dependencies. As the number of dependencies increase, so does the likelihood that a method that the developer would like to offload depends on another method or class which contains a trouble spot, which, in turn prevents the offloading. It is possible to partly overcome this problem by creating smaller classes having fewer dependencies and internal state. 

Our experiences clearly suggest that manually identifying the methods having offloading problems is non-trivial. It may include many cycles of trial-and-error and the only way to know whether all issues have been resolved is to test the application to see whether it works correctly or not. We think that it is essential to have tools that can automate this procedure and guide the programmers into developing more offloadable code.

\subsection{Energy consumption of local vs. offloaded Twitter}
\label{sec:andtweet_energy}

After eventually offloading Twitter communication, we measured the energy
consumption of non-modified and offloaded AndTweet. We tested offloading under two different network
conditions. In first setup the offloading server is in the same Wi-Fi
network as the phone. In the second scenario, the phone used 3G as the access network.

The main energy
consumers in mobile phones are CPU, I/O, display, and network interfaces. 
We used the Monsoon Power
Monitor (www.msoon.com)
to measure the energy consumption of the phone.  We also
collected the packet traces and used models to estimate the
energy consumed by the network interfaces. We used the model presented in \cite{yu10eenergy} to
estimate Wi-Fi energy consumption. As for 3G, we use a deterministic
power model which takes into account the different operating modes and
state transition controlled by inactivity timers according to the 3G
Radio Resource Control protocol\cite{balasubramanian09imc}. Power draw
of each state and the inactivity timer values were measured
beforehand. Based on the timer values and the observed traffic
patterns, the model deduces the time spent in each 3G radio state and
computes the energy consumption estimates. In the measurements and
the estimates, we excluded the one-time cost of transferring the
application image, because the applications could be, for example,
pre-installed to the offloading infrastructure.

\begin{table}[t]
\begin{scriptsize}
\begin{tabular}{ l l l }
   Measurements & Wi-Fi(avg/stdev) & 3G(avg/stdev) \\
  \hline
  Total energy & 2.67/0.59 J & 3.92/1.97 J \\
  Total en., offloaded & 25\% less/0.3 J & 18\% more/2.1 J \\
  Network en., offloaded & 46\% less/0.04 J & 33\% more/2.0 J \\
  Execution time & 2.69/0.59 s & 3.86/2.02 s\\
  Exec. time, offloaded & 6\% more/0.5 s & 33\% more/2.1 s\\
  Traffic & 7.7/0.8 kB & 6.0/2.1 kB  \\
  Traffic, offloaded & 17\% less/0.5 kB & 31\% less/1.8 kB \\ 
\end{tabular}
\caption{AndTweet offloading measurements for single refresh event. Total is measured and network is estimated energy consumption.}
\label{tab:andtweet-reload}
\end{scriptsize}
\end{table}

The results in Table \ref{tab:andtweet-reload} show energy consumption
of a single Twitter event which checks and fetches new tweets, and
then displays them. We
observe that offloading saves one fourth of the total energy consumed in the
Wi-Fi setup. As expected, the savings clearly come from having less
network traffic. However, the execution takes slightly longer
when offloaded, which increases the energy consumed by the
display. The results are very different when switching to 3G. More energy is consumed with the offloaded version even if less data
is transmitted and received. Because the RTT in 3G is an order of magnitude longer than with Wi-Fi, the remote invocation takes more time. Combined with long 3G inactivity timer values, this causes the network interface to be in
active high-power state (DCH) during the whole invocation. Execution time also has a big variance, because 3G latencies vary depending on network conditions and the current state of the radio when starting the transmission. Obviously, both network conditions and traffic patterns play a major role in the profitability of
offloading.

\section{SmartDiet}
\label{sec:toolkit}

\subsection{Overview}

In view of the experiences described above, we argue that the software
developer must be included in the offloading process. Furthermore,
a set of analysis tools are necessary to make this process
feasible. We therefore propose such a toolkit which we envision to comprise three tools: energy
analysis, constraint identification and software structure
analysis. We have designed and implemented prototypes of the
first two, but currently have only a vision of the third one.

The energy analysis tool finds and visualizes the parts of a
given application that could yield energy savings when offloaded. The
constraint identification tool automatically identifies constraints
in the source code, determines which methods can be offloaded as
such and points out trouble spots in the code. The idea for the
software structure analysis tool is that it analyzes the program code
to find opportunities to save energy through restructuring, for example by merging methods or classes. 
The developer would first use the energy analysis tool to identify
the candidates for offloading. Next, the constraint identification
tool would be used to find and resolve problems in those candidates. Finally,
a structure analysis tool could be used to explore ways of making further
energy savings. In case such opportunities were found, the first two
tools would be applied again. This process will result in an application
that can be offloaded with larger energy savings by employing existing
frameworks.

\subsection{Energy analysis tool}
\label{sec:energy-usage-analysis-tool}

\subsubsection{Collecting measurements}

The energy analysis tool implements a measuring and modeling
setup for profiling and visualizing the energy consumption of a
given application. Since we collect data dynamically while the
program is running, the results reflect the specific usage scenario that is run. 
The tool needs to collect three kinds of information:
the amount of computation, the amount of communication, and a trace of the
program execution flow to later produce class and method level
statistics for the developer.

Our tool collects information about the communication by capturing all IP traffic in the device, annotated with timestamps. To implement this, we
inject a kernel module using netfilter hooks
(www.netfilter.org) to access this data in
real-time. To collect CPU usage statistics, the tool uses
oprofiler (http://oprofile.sourceforge.net) in order
to do similar HPC-based computational energy profiling as in
\cite{yu10hpc} or analyze the /proc
filesystem as in
\cite{zhang10powertutor}. Both oprofiler and /proc filesystem
polling are in their unmodified form unable to gather data in a
sufficiently fine-grained fashion without adding a major overhead to the
overall system performance. In order to improve its performance, we plan to customize oprofiler to collect only the data we need. To track the program execution, the tool uses execution tracking
features in Android Debug Monitor Server
(DDMS)
which produces a trace of all classes and methods executed during the
run. In addition, we implemented minor modifications to the Dalvik
virtual machine to annotate the traces with system-wide timestamps, which
can be matched to those of our other measurements.

\begin{figure}[t]
\centering
\includegraphics[width=240pt, height=150pt]{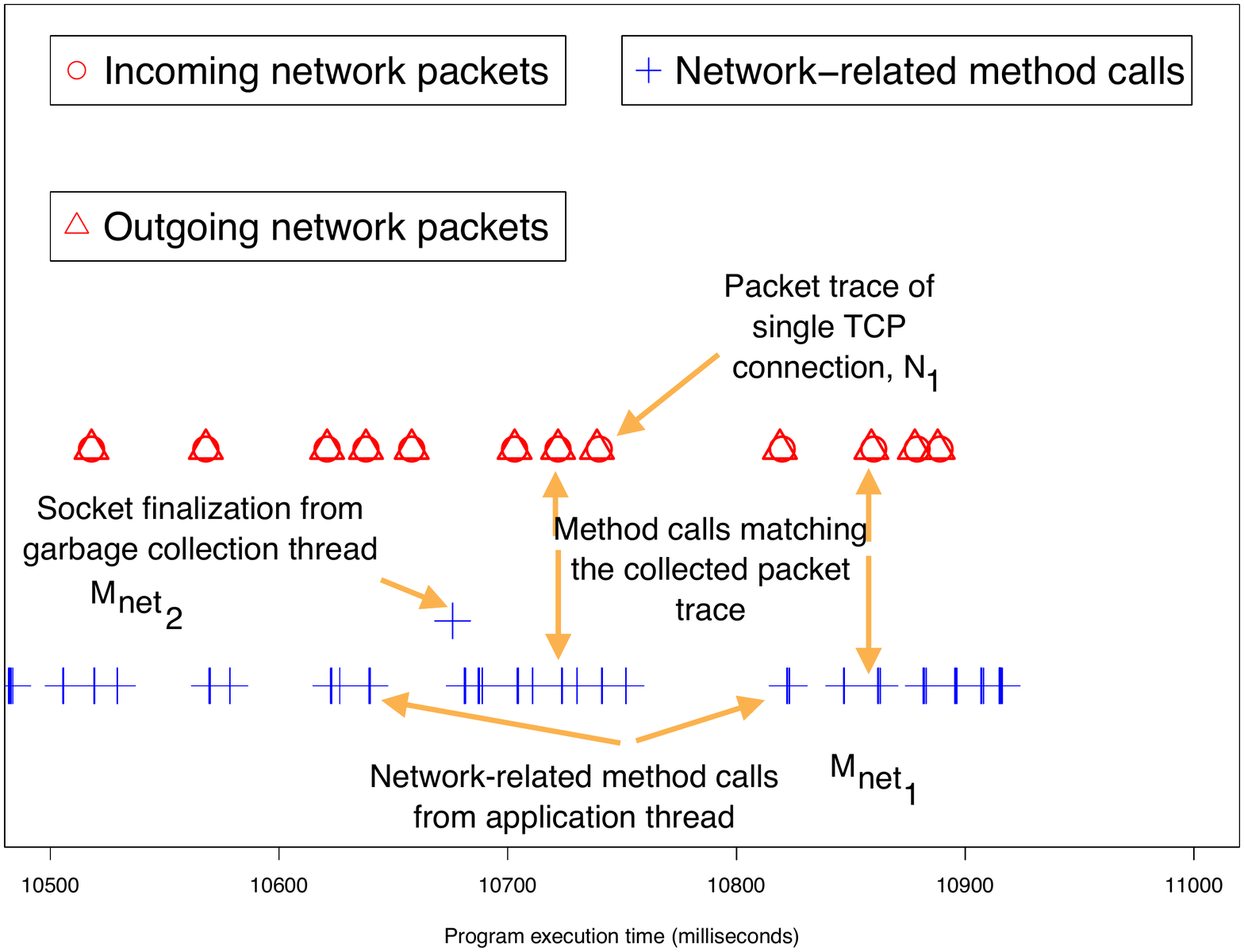}
\caption{TCP packets and network-related method calls for test application fetching an HTML page.}
\label{fig:methods-and-packets}
\end{figure}

\begin{figure*}[t]
\centering
\includegraphics[width=500pt]{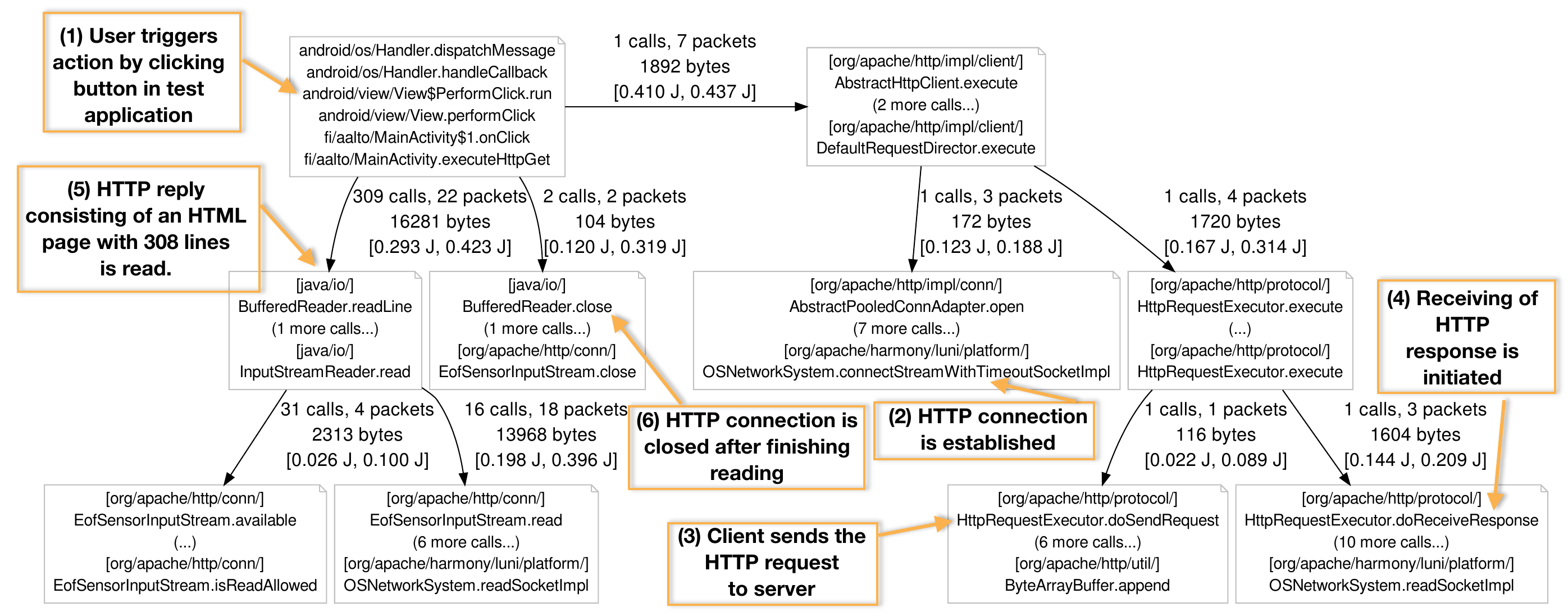}
\caption{Network usage graph for test application, which fetches an HTML page. Numbers show how many times methods have been invoked during the whole procedure and how much energy it consumed.}
\label{fig:simpleapp}
\end{figure*}

\subsubsection{Matching packets to method trace}

SmartDiet associates each packet in the collected packet trace
to an individual method in the program execution trace. 
The tool starts by dividing the execution trace into threads
and packet trace into separate flows (TCP connection or UDP
flow). Furthermore, only network-related method calls are
filtered for each thread. We now have two separate time series:
network-related method calls of each thread and packet arrival events of each
flow (see Figure \ref{fig:methods-and-packets} for
illustration). These series are compared by computing
cross-correlations in order to associate each flow to a particular
thread which is generating that traffic. The idea is that each
network-related method is associated with the corresponding
packets. Finally, each packet of a flow is associated to the closest
(in time) method call of the corresponding thread. This way, the
tool generates a method trace of the program execution annotated with
information about the methods that caused network traffic.

Figure \ref{fig:methods-and-packets} shows part of the traces of a
simple Android test application that performs an HTTP GET request when
a button is clicked. The thread executing the HTTP request correlates
strongly with the packet trace. Another thread in the figure has a single
network-related call. It is the garbage collector thread
running finalization for a network-related object that is no longer
used. Since it correlates weakly with the packet trace, no packets are
associated with it.

\subsubsection{Visualizing the network usage}

Program execution for each thread can be viewed as a hierarchical call
tree, where a method calls another method which calls another and so
on. Our tool reconstructs this tree, carrying along the information of
the detected network usage. It then aggregates the traffic of the
nodes up in the tree, so that the root method, where the execution
starts, gets associated with all packets that have been sent or
received within each thread.

Figure \ref{fig:simpleapp} is an example graph, automatically produced by SmartDiet, for a simple test case requesting an HTML document over HTTP. Traffic is cumulatively assigned to the MainActivity.onClick
method and from there on, divided between various library functions
that open the connection, send an HTTP request, receive the response
and finally close the connection. At each step, we present the number
of calls made and traffic statistics alongside with
energy consumption estimates based on the models described in
Section \ref{sec:andtweet_energy}.

The energy usage estimate is shown as a range between two values,
because in estimation one has to make assumptions about the
dependence of traffic in a single method to the traffic of the rest
of the program. Let $N$ be the full packet trace,
$N_{method}$ the part of the packet trace associated with a given
method and $N_{rest}$ everything from $N$ that does not belong to
$N_{method}$. $E(n)$ is the energy-estimate for a packet trace $n$ and hence $E(N)$ energy-estimate for the whole program.

Assuming that all other traffic is left unmodified as we offload the
traffic of a single method, $E(N_{rest})$ would be the remaining
energy consumption. The energy savings would hence be $E(N) -
E(N_{rest})$. If we assume that all other traffic is independent of
our method, we can calculate the energy saved as $E(N_{method})$. The latter estimate is always larger than
the former, because sending multiple packets together is always
energy-wise cheaper than sending them separately. Because we do not estimate the dependencies between traffic caused by different methods, we take the
former as the lower bound of energy savings $E_{min}$ and the latter
as the upper bound $E_{max}$, and let the developer do further reasoning.

When we tried this approach on more complex 
applications, mainly AndTweet and
ConnectBot\footnote{\url{http://code.google.com/p/connectbot/}}, we identified
a caveat: Using the method trace information only, we are not always able to accurately track the data flow between threads. If an application uses
a threading network library--the case especially with
ConnectBot--our aggregated results will show energy consumption only within
the library code. The developer, however, is mostly interested in
those parts of the application code that actually generated this
traffic. Section \ref{sec:challenges} discusses potential solutions to this problem of execution tracking.

\subsection{Constraint identification tool}
\label{sec:constraint-identification-tool}

SmartDiet's constraint identification performs the analysis on the application source code. For each method in the application, it points out problems that can prevent offloading of that method unless the code is modified. SmartDiet currently focuses on heuristics that identify problems associated with our offloading setup, which is using the Android platform and Java Serialization API to implement the remote execution of methods. Nevertheless, similar heuristics can be crafted to other remote execution mechanisms, including the Android Parcelable mechanism used in Cuckoo \cite{kemp10cuckoo}, .NET serialization used in MAUI\cite{cuervo10maui}, or even the virtual machine based thread migration used in CloneCloud \cite{chun11clonecloud}, because they all set some restrictions on what kind of methods can be offloaded. 

\subsubsection{Identifying hardware constraints: access to local resources}

The first set of constrained methods are those that require access to the hardware of the local device. If a method accesses one of the constrained Android system APIs, it cannot be offloaded unless the code structure is changed. We currently identify method as having this constraint if it tries to show, for instance, notifications to the user, update anything on the screen, vibrate the phone, access the Bluetooth, wifi or usb subsystem, and so on. We have identified a total set of 20 constrained subsystems.

\subsubsection{Identifying software constraints: unexpected behavior or limitations in migratability}

Our second set of constrained methods are those that cannot be migrated to the remote server at all due to migration mechanism requirements, or those that cause unexpected behavior when executed remotely due to inconsistent states between local and remote execution environments.

Migration limitations are specific to the mechanism used, which in our case is the Java serialization APIs. For a method to be migratable, its encapsulating class as well as arguments and return type must implement the Java serializable interface. SmartDiet finds all methods that adhere to this criterion to show which ones can be directly migrated. 

We find that only a fraction of analyzed methods implement the serializable requirements. We therefore also calculate the number of methods which could be modified to be serializable with a few minor changes. In this category, we include all methods whose encapsulating class, as well as arguments and return types, are convertible to serializable with the following criteria: A type is considered convertible to serializable, if all of its supertypes and member classes are either directly serializable or belong to this application's codebase and can be also converted to serializable using this same principle recursively. We exclude the library code, because for instance the Android SDK code cannot usually be easily modified, even if the changes would be simple.

Regarding unexpected behavior, SmartDiet finds all methods that access the local file system using either Android's SharedPreferences mechanism or Java's File class. ThinkAir does not synchronize the file system, thus files in the remote server are not the same as those in the device, which will often cause unexpected behavior for the program. This principle can be extended to find problems related to other non-synchronized resources as well.

A potential solution to the synchronization issues is a system which  automatically synchronizes the relevant state. This is notably a hard problem on its own, but in the context of offloading we are also concerned about energy usage of the synchronization. Until efficient automatic solutions are presented, the developer must be assisted in overcoming the problems manually.

Based on our experiences with AndTweet, we believe that an even richer
set of rules could be established to detect and classify different
types of remote execution issues, as well as provide suggestions to the
developer for overcoming them. For example, if a class contains
an instance of the non-serializable HttpClient class, as described in
Section \ref{sec:constraints-of-remote-execution}, we might suggest
removing the member instance and replacing it with a new instance of
HttpClient created on-the-fly every time it is needed.

\subsubsection{Statistics from open source programs}
\label{sec:program-statistics}

\begin{table}
\begin{scriptsize}
\begin{tabular}{ p{4cm} l l l l }
Statistic & Median & Min & Max \\
\hline 
Number of methods & 431 & 121 & 4411 \\
Directly migratable & 0.17\% & 0.00\% & 3.70\% \\
Migratable with minor changes & 15.7\% & 0.00\% & 46.8\% \\
Hardware access constraints & 14.2\% & 2.28\% & 41.3\% \\
Potential unexpected behavior & 10.7\% & 0.00\% & 30.3\% \\
because of access to file system & & & \\
\end{tabular}
\end{scriptsize}
\caption{Constraint statistics for 16 open source applications.}
\label{tab:heuristics}
\end{table}

Our tool analyzes source code, which restricts us to analyzing
open-source software. Unfortunately the most popular applications on
Android Market are closed-source. To find similar applications, we
went through numerous Android application listings, and selected any programs that are
non-trivial in size and include either communication or non-trivial
computation\footnote{We used Android Market, popular source code sharing site Github, Wikipedia and numerous other sources}. Additionally we analyze also some platform applications
that are shipped with Android operating system, like the web browser.

We ran our constraint analysis tool on this set of programs. Results are shown in Table \ref{tab:heuristics}. Maximum of 3\% of methods are directly migratable, but SmartDiet can point out changes to source code which will enable the migration of 15 to 47\% of methods for offloading using ThinkAir. SmartDiet can also guide the developer into fixing the issues regarding hardware or filesystem access. 

However, even after these changes, an overwhelming majority of methods have migration issues. Here's where the third tool in our toolset, one that performs structural analysis, would be helpful. We could guide the developer in making larger structural changes, which would enable new portions of the application to be offloaded, in an energy-efficient fashion.


\section{Remaining challenges}
\label{sec:challenges}

In our current prototype we have implemented dynamic measurement of communication energy usage. Existing solutions can be applied to estimate computation and display energy consumption\cite{yu10hpc,zhang10powertutor} given just that time stamping and high enough sampling rates can be supported. We plan to integrate CPU usage measurement and estimation to the same toolkit. We can find CPU hot spots in a similar fashion as we have done for network transmission. 

As discussed in Section \ref{sec:energy-usage-analysis-tool}, we encountered problems when analyzing the execution flow of complex programs, relying heavily on threading. We are currently looking into the TaintDroid \cite{enck10taintdroid} system, which modifies the Android Java virtual machine in such a way that data flows can be tracked, and would like to extend our toolkit using similar mechanisms for data and execution flow tracking over thread boundaries. 

Regarding our network energy usage modeling, we currently use only the traffic as input. In order to explicitly show the impact of network conditions, the power models can be improved by using metrics reflecting network conditions as parameters.

Conserning the application structure analysis tool, we believe it would be beneficial to investigate which kind of programming styles and application structures best suit offloading. Combining the results of our heuristics to standard object-oriented code quality measures, like coupling and cohesion, could yield interesting results and insights into how we can enhance the effectiveness of offloading. This can also lead to suggestions how to improve programs regarding offloadability. 

\section{Conclusions}

In this paper we studied feasibility and potential energy savings that are achievable utilizing method-level offloading, especially in network-intensive mobile applications. We used an open source Twitter client as an example to show that completely automated offloading often misses opportunities for saving energy and may also lead to various failures during execution. For this reason, we argue that the developer should participate actively in the offloading process. To this end, we propose an offloading analysis toolkit called SmartDiet. It analyzes Android application source code and collects run-time information in order to help developers in identifying potential energy savings and trouble spots in their program code so that existing offloading frameworks can be more efficiently utilized. SmartDiet has shown promising results in helping the development of offloadable code, although some challenges still remain to complete all the features we believe would be useful.

\bibliographystyle{abbrv}
\bibliography{offloading}  

\end{document}